\documentclass[%
 reprint,
 amsmath,amssymb,
 aps,
]{revtex4-1}

\usepackage{graphicx}
\usepackage{dcolumn}
\usepackage{bm}

\begin{document}
\bibliographystyle{apsrev4-1}

\preprint{APS/123-QED}

\title{Neutron diffraction study and anomalous negative thermal expansion 
in non-superconducting PrFe$_{1-x}$Ru$_x$AsO}

\author{Yuen Yiu$^a$, V. Ovidiu Garlea$^b$, Michael A. McGuire$^c$, Ashfia Huq$^d$, David Mandrus$^{a, c, e}$ and Stephen E. Nagler$^{b, f}$
}
\affiliation{
$^a$ Department of Physics and Astronomy, University of Tennessee;
$^b$ Quantum Condensed Matter Division, Oak Ridge National Laboratory; 
$^c$ Materials Science and Technology Division, Oak Ridge National Laboratory; 
$^d$ Chemical and Engineering Materials Division, Neutron Sciences Directorate,  Oak Ridge National Laboratory;
$^e$ Materials Science and Engineering, University of Tennessee; 
$^f$CIRE, University of Tennessee.
}%

\date{\today}

\begin{abstract}
Neutron powder diffraction has been used to investigate the structural and magnetic behavior of the isoelectronically doped Fe pnictide material PrFe$_{1-x}$Ru$_x$AsO.  Substitution of Ru for Fe suppresses the structural and magnetic phase transitions that occur in the undoped compound PrFeAsO. Contrary to the behavior usually observed in 1111 pnictide materials, the suppression of both the structural and magnetic transitions does not result in the emergence of superconductivity or any other new ground state.    Interestingly, PrFeAsO itself shows an unusual negative thermal expansion (NTE) along the c-axis, from 60K down to at least 4K; this does not occur in superconducting samples such as those formed by doping with fluorine on the oxygen site.  We find that NTE is present for all concentrations of PrFe$_{1-x}$Ru$_x$AsO with x ranging from 0.05 to 0.75.  These results suggest that the absence of   superconductivity in these materials could be related to the presence of NTE.

\end{abstract}

\pacs{Valid PACS appear here}
\maketitle


\section{\label{sec:level1}Introduction}

Since the discovery of Fe-based superconductors in 2008\cite{kamihara08, lynn09} there has been enormous interest in obtaining a detailed understanding of the properties of closely related materials.  Studies of substitutionally altered or doped materials have proved particularly valuable and have led to the discovery of several superconducting families.  Here we report new results for the isoelectronically doped 1111 pnictide material PrFe$_{1-x}$Ru$_x$AsO.

The 1111 FeAs based materials (typified by LaFeAsO) share several common characteristics with most of the other Fe-based superconductors. As the temperature is lowered the undoped parent material does not reach a superconducting ground state, but undergoes a structural transition, followed or accompanied by an antiferromagnetic spin density wave (SDW) transition of the moments associated with the Fe atoms and sometimes antiferromagnetic (AFM) ordering of the rare earth moments\cite{lynn09}.  For superconductivity to emerge in the material, all of these transitions must be eliminated or suppressed down to a sufficiently low temperature\cite{canfield09, sanna10}.  This can be done by doping with the appropriate element, e.g. F on the O-site in CeFeAsO\cite{zhao08}, Co on the Fe-site in LaFeAsO\cite{sefat08, wang09}, etc.  In certain members of the Fe oxypnictides, superconductivity can also be induced by simply applying pressure externally, e.g. LaFeAsO under 120kbar for a T$_C$ of 21K\cite{okada08} and BaFe$_2$As$_2$ under 40kbar for a T$_C$ of 29K\cite{alireza09}.  The application of pressure does not always produce this result, for example CeFeAsO achieved no superconductivity under pressures up to 500kbar\cite{zocco11}.

There are situations where AFM ordering and superconductivity can coexist, for example in 122 materials such as in BaFe$_{2-x}$Co$_x$As$_2$\cite{nandi10}.  Although this occurs in several 122 materials \cite{lumsden10} the general presumption that superconductivity and AFM ordering are competing ground states remains valid\cite{kim11, tanabe11}.  Notably coexistence is much less common in 1111 materials such as LaFeAsO, where long range AFM ordering is usually completely destroyed before superconductivity can emerge, although there are some exceptions such as SmFeAsO$_x$F$_{1-x}$\cite{drew09}. The suppression of the structural and magnetic transitions can also lead to new non-superconducting ground states, for example ferromagnetism\cite{luo10}.  It has been concluded that suppressing the structural and magnetic transitions is usually necessary, but not sufficient, for inducing superconductivity\cite{canfield09}. 

\begin{figure}
\includegraphics[width=65mm]{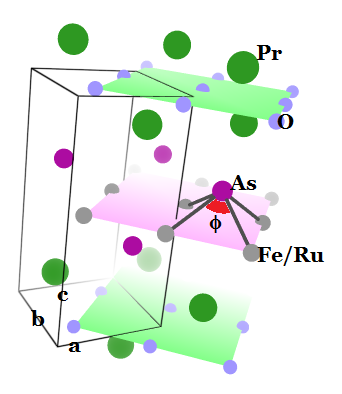}
\caption{The PrFe$_{1-x}$Ru$_x$AsO room temperature tetragonal unit cell showing the rare-earth oxygen (top, bottom) and Fe-As (middle) layers, the Fe-As pyramid and the Fe-As-Fe angle $\phi$.  Note that in the orthorhombic structure there are two inequivalent values for $\phi$.}
\end{figure}

The unit cell of the 1111 parent compound PrFeAsO is illustrated in Figure 1.  Most of the materials with the highest reported superconducting T$_C$'s are doped on either site of the rare-earth-oxygen layer\cite{lynn09, tanabe11}.  Superconducting T$_C$'s are generally lower for materials doped on the Fe-As layer, possibly due to undesirable lattice distortions and disorder in the conducting Fe-As layer\cite{sharma10, delacruz10}.  To better understand this, isoelectronic substitution on the Fe site has proved very instructive since unlike carrier doping it enables the study of materials without directly altering the Fermi level\cite{mcguire09a}.  Moreover, isoelectronic doping with Ru can lead to superconductivity in certain Fe-oxypnictides, such as the 122 materials, BaFe$_{2-x}$Ru$_x$As$_2$\cite{sharma10} and SrFe$_{2-x}$Ru$_x$As$_2$\cite{qi09}.  In contrast, however, Ru/Fe substitution in 1111 materials has not been observed to lead to superconductivity even after the complete suppression of the structural and magnetic transitions: the studied systems include La(Fe,Ru)AsO\cite{sanna10}, Pr(Fe,Ru)AsO\cite{mcguire09a}, Sm(Fe,Ru)AsO and Gd(Fe,Ru)AsO\cite{pal10}.

McGuire \textit{et al.} showed that Ru doping in PrFeAsO does not induce superconductivity down to at least T = 2K\cite{mcguire09a}.  Even when the structural and magnetic transitions were suppressed completely there was no indication of any new ground state\cite{mcguire09a}.  Bulk and transport measurements on PrFe$_{1-x}$Ru$_x$AsO showed hints of anomalies associated with a possible phase transition for concentrations up to x = 0.5 or 0.6\cite{mcguire09a}.  This is a rather high level when compared with carrier doped Fe-pnictides, where the transitions are suppressed with less than 10\% doping in most cases.  Thus, Ru-doping provides a wide stoichiometric window for studying the progressive suppression of the structural and magnetic transitions.   Ru-doping studies have been also carried out on F-doped superconducting samples, such as LaFe$_{1-x}$Ru$_x$AsO$_{0.89}$F$_{0.11}$\cite{satomi10} and SmFe$_{1-x}$Ru$_x$AsO$_{0.85}$F$_{0.15}$\cite{tropeano10}.  The general results show that the superconductivity in these samples is not sensitive to isoelectronic doping except at very high levels of the order of x = 0.3 and above. 

A phenomenon of particular interest in the  parent compound PrFeAsO is the observation of negative thermal expansion (NTE) along the c-axis, where the sample gradually expands along the c direction as it is cooled below 60K\cite{kimber08}.  Notably such NTE is absent in superconducting F-doped PrFeAsO\cite{kimber08}.  Recent results have shown that NTE also occurs in NpFeAsO were it is associated with magnetic ordering of the Np ions\cite{klimczuk12}.  A different NTE behavior has been observed in the superconducting samples of Co doped BaFe$_2$As$_2$ where a sudden onset of NTE along the c axis appears at T$_C$.  This feature is absent in the non-superconducting parent BaFe$_2$As$_2$\cite{daluz09}.  It seems worthwhile to explore whether or not there is a systematic correlation between the NTE and the presence or absence of superconductivity. Despite voluminous reports of the dependence of lattice constants on stoichiometry in Fe-As based materials there has been surprisingly little published data showing the temperature dependence of the lattice constants, especially the c-axis.   
 
In this paper we report new results on Ru doped PrFe$_{1-x}$Ru$_x$AsO, , including detailed studies of the temperature dependence of lattice parameters.  As mentioned above undoped PrFeAsO shows NTE.  As discussed below, we observe that Ru-doping in PrFeAsO suppresses the structural and magnetic phase transitions without leading to superconductivity; however NTE along the c axis persists for all doping levels up to at least 75\% Ru on the Fe site. This paper is organized as follows: Sample synthesis and preliminary characterization is described in section II, the results of neutron scattering experiments are described and discussed in section III, with conclusions in section IV.

\section{\label{sec:level2}Sample synthesis and characterization}

\begin{figure}
\includegraphics[width=70mm]{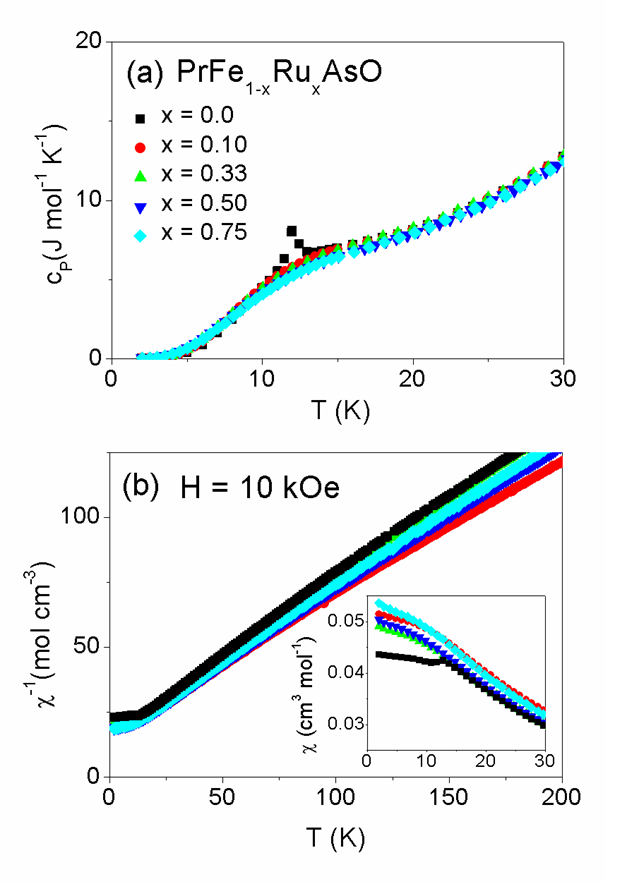}
\caption{(a): Heat capacity vs T for PrFe$_{1-x}$Ru$_x$AsO, the peak at 14K indicates the AFM ordering of Pr in PrFeAsO. No peak is observed for x $\ge$ 0.1; 
(b):$\chi$$^{-1}$ vs T, the inset shows a closer look at the lower temperature range for $\chi$ vs T.}
\end{figure}

Polycrystalline samples of PrFe$_{1-x}$Ru$_x$AsO with x = 0.05, 0.1, 0.33, 0.4, 0.5, 0.6 and 0.75 were synthesized using methods reported previously by McGuire \textit{et al.}\cite{mcguire09a}.  In each case, a stoichiometric mixture of powders of PrAs, Fe$_2$O$_3$, RuO$_2$, Fe and Ru was crushed and mixed thoroughly in a glove box, pressed into pellets and received two firings at 1200-1250$^o$C.  Transport property data and results of electronic density of state calculations for the same series of materials have been reported previously\cite{mcguire09a}.  
The undoped parent compound PrFeAsO undergoes a tetragonal-orthorhombic structural transition at T $\sim$ 150 K, followed by a SDW transition at T $\sim$ 140 K associated with the AFM ordering of Fe, and then finally AFM ordering of the Pr moments at 14 K\cite{mcguire09b}.  Previously published transport measurements for PrFe$_{1-x}$Ru$_x$AsO have shown evidence for the suppression of long range order of the Pr moments for x $\ge$ 0.1 and structural/SDW transitions for x ${ >  0.5}$\cite{mcguire09a}.

Prior to the neutron powder diffraction experiments, we have carried out heat capacity and magnetization measurements.  As seen in Fig. 2(a) and (b), the AFM ordering of Pr at 14K in the x=0 material is indicated by a peak in the heat capacity and a kink in the magnetization.  These features disappear for x $\ge$ 0.1.

\section{\label{sec:level3}Experimental results and discussion}

\begin{figure}
\includegraphics[width=90mm]{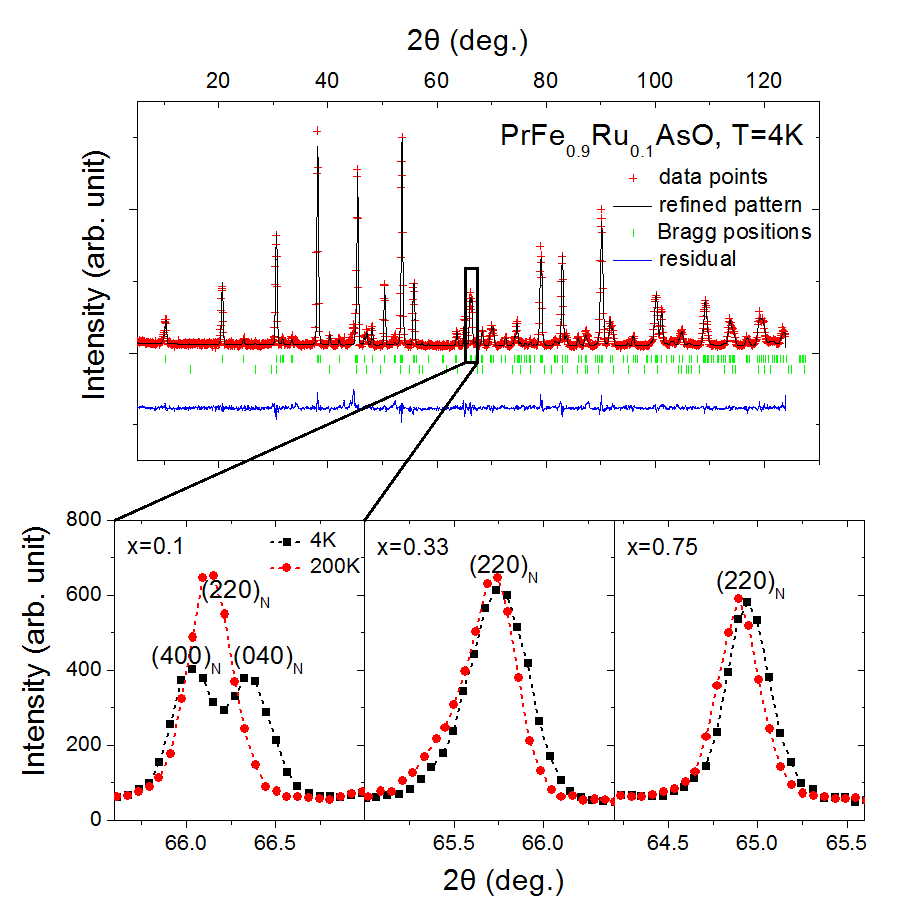}
\caption{(Upper): Fully refined neutron powder diffraction pattern collected at HB2A of HFIR for PrFe$_{1-x}$Ru$_x$AsO at 4K; (Lower): 220$_N$ peak for x = 0.1, 0.33 and 0.75 at T = 200K (red circles) and 4K (black squares).  Peak splitting is clearly visible only for x=0.1.} 
\end{figure}

\begin{figure}
\includegraphics[width=90mm]{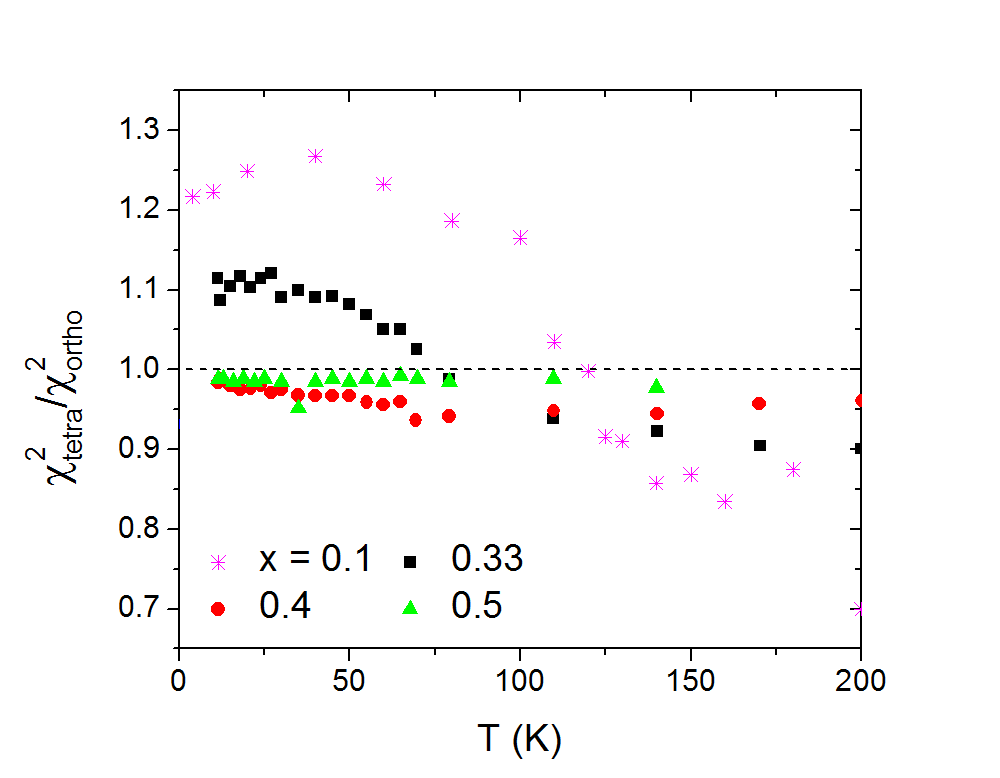}
\caption{The ratio $\chi^2_{tetra}$ given by the Rietveld refinement with the P4/nmm (tetragonal, space group 129) structure, to $\chi^2_{ortho}$ obtained from refining to the Cmma (orthorhombic, space group 67) structure. Any value larger than 1.0 indicates the Cmma model provides a better fit than P4/nmm.  The represents data collected from both HB-2A and POWGEN.} 
\end{figure}

\begin{figure}
\includegraphics[width=80mm]{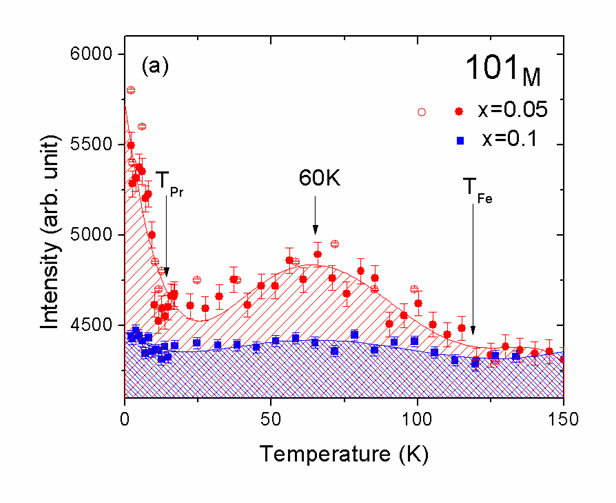}
\includegraphics[width=83mm]{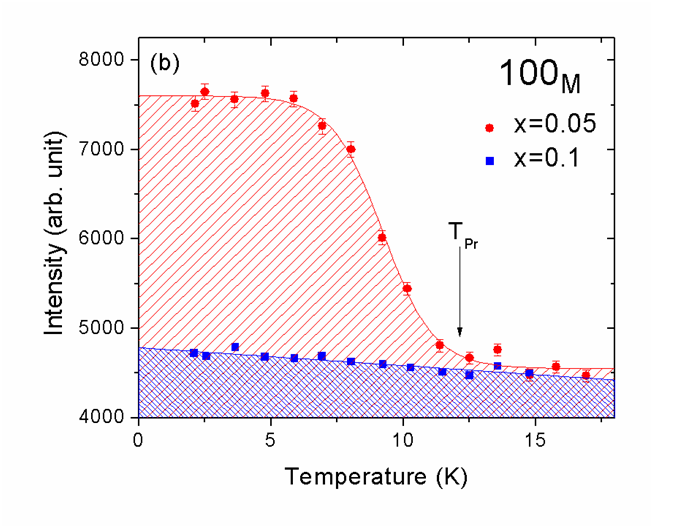}
\caption{(a): The intensity of the 101$_M$ peak as a function of temperature.  The AFM ordering of both Pr and Fe are visible in x = 0.05 but not x = 0.1.  The intensity of the 101$_M$ peak reaches a domelike maximum around 60K for x = 0.05.  Solid dots represent single data points collected from counting at the center peak position; hollow dots represent normalized data from fits of entire magnetic peaks.  The lines and shadings are guides to the eye; 
(b): Intensity of the 100$_M$ peak as a function of temperature, showing the size difference of the Pr moment between x = 0.05 and 0.1.}
\end{figure}
\begin{figure}
\includegraphics[width=85mm]{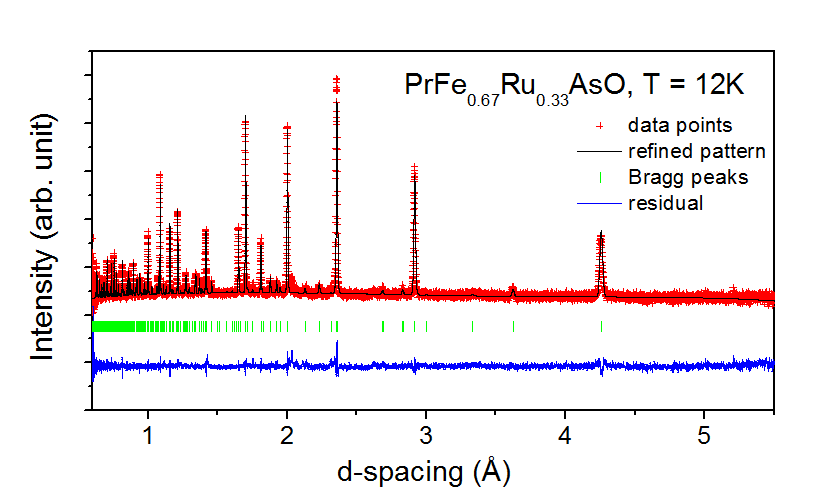}
\caption{Neutron powder diffraction data collected from POWGEN refined to the orthorhombic Cmma structure for x = 0.33 at T = 12K.}
\end{figure}

\begin{table*}
\caption{Structural parameters and $\chi$$^2$ values from Rietveld refinement of NPD data from HB2A (for Ru\% = 10 only) \\and POWGEN, at T = 12K.
(Space group: Cmma (\#67) for 10\% and 33\%, P4/nmm (\#129) otherwise.)}

\begin{ruledtabular}
\begin{tabular}{ccccccc}

\textrm
Ru\%  & a/b (\AA)				&c (\AA)		&z$_{As}$		&z$_{Pr}$		&$\phi$	&$\chi$ $^2$\\
\colrule
10	&3.96865(8)/3.98509(8)		&8.5622(2)   &0.6550(5)  	&0.1402(7) 	&71.89(9)/72.23(9)	&3.34	\\
33	&4.00063(3)/4.00733(3)		&8.5154(1)	&0.6570(3)	&0.1389(5)	&71.95(7)/72.09(7)	&3.64	\\
40	&4.01162(4)				&8.4911(2)	&0.6582(5)	&0.1378(6)	&71.97(7)	&3.51	\\
50	&4.02491(6)				&8.4613(2)	&0.6626(2)	&0.1328(5)	&71.43(7)	&2.82	\\
60	&4.03351(7)				&8.4364(3)	&0.6603(6)	&0.1333(9)	&71.93(6)	&4.04	\\
75	&4.05088(4)				&8.3980(1)	&0.6587(4)	&0.1366(6)	&72.41(7)	&3.26	\\

\end{tabular}
\end{ruledtabular}
\end{table*}
	We performed neutron powder diffraction on PrFe$_{1-x}$Ru$_x$AsO using the neutron powder diffractometer HB-2A (for x = 0.1, 0.33 and 0.75) and the fixed incident-energy triple-axis HB-1A (for x = 0.05 and 0.1) of HFIR at ORNL to study the structural and magnetic transitions respectively.  Additional measurements were performed at the high-resolution powder diffractometer POWGEN of SNS at ORNL to study the effect of Ru doping on the anomalous NTE and also to have a closer look at the suppression of the structural transition (for x = 0.33, 0.4, 0.5, 0.6 and 0.75).

	The detailed structures were fitted via Rietveld refinement using the FULLPROF program\cite{caravajal93}.  Fig. 3 (upper) presents the refined neutron powder diffraction data collected at HB-2A for the x = 0.1 sample at the base temperature of 4K. 	Fig. 3 (lower) shows a closer look at the evolution of the 220 nuclear peak for x = 0.1, 0.33 and 0.75. The structural transition is evident in the x = 0.1 sample as the peak splitting in 220$_N$ is easily discernible; this is much less obvious for samples with x $\ge$ 0.33.  Fig. 4 shows the temperature dependence of the relative goodness of fits obtained by fitting the full powder patterns to each of the two different known structural models (tetragonal P4/nmm vs. orthorhombic Cmma.)   It is apparent that the orthorhombic model fits better than the tetragonal model for x = 0.1 at T $\le$ 120K and for x = 0.33 at T $\le$ 75K.  The inferred structural phase transition temperature in the x = 0.33 sample is consistent with the anomaly near 80K observed in the resistivity data\cite{mcguire09a}.  The diffraction measurements for higher Ru concentrations provide no evidence for any structural transition, suggesting that it is suppressed for Ru doping between 33-40\%.

	Interestingly for other isoelectronically doped 1111 materials the structural phase transition has been observed to persist to dopings well above the 10\% level;  for example in CeFeAs$_{1-x}$P$_x$O  both the transition to an orthorhombic structure and antiferromagnetic order are present up to x = 0.4\cite{luo10}.   In contrast, a much smaller level of carrier doping is needed to affect the transitions. In RFeAsO$_{1-x}$F$_x$ all structural and magnetic transitions were suppressed with less than 10\% doping\cite{kamihara08, rotundu09, ren08a, ren08b, ren08c}.  In fact the doping level resulting in the highest observed superconducting transition temperature occurs at x = 0.1 for R = La\cite{kamihara08}, 0.11 for Pr\cite{rotundu09, ren08a}, 0.11 for Nd\cite{ren08b} and 0.1 for Sm\cite{ren08c}, with superconductivity usually disappearing around x = 0.2\cite{ren08a, ren08b, ren08c}.

	  We now turn to a detailed examination of the magnetic transition that is observed in PrFe$_{1-x}$Ru$_x$AsO with 5\% Ru doping.  In Fig. 5 we present temperature dependent measurements of the intensities of the 101$_M$ and 100$_M$ magnetic peaks measured using HB-1A for doping levels x = 0.05 and 0.1.  The 101$_M$ peak shows contributions from both Pr and Fe moments while the 100$_M$ peak is sensitive only to the Pr moments.  It is apparent in Fig. 5(a) and (b) that magnetic ordering of both Pr and Fe exists at x = 0.05 but is suppressed at x = 0.1.  Fe orders antiferromagnetically in the x = 0.05 sample at 120K which may be compared to 137K\cite{mcguire09b} observed in the undoped PrFeAsO; no AFM order is visible in x = 0.1.    Evidently in PrFe$_{1-x}$Ru$_x$AsO, the magnetic transition is more sensitive to Ru doping than is the structural transition.  In any case neither transition is observed for x $\ge$ 0.4.

	The temperature dependence of the magnetic intensity of the 101$_M$ peak is particularly interesting. As seen in Fig 5(a), in the x = 0.05 sample the intensity of the 101$_M$ peak reaches a dome-like maximum at T $\sim$ 60 K, followed by a modest decrease as the temperature is lowered, increasing again below T $\sim$ 14 K, the AFM ordering temperature for Pr.    This feature has also been observed in undoped PrFeAsO at T $\sim$ 60K\cite{kimber08}.  Critical fluctuations in the Pr magnetic subsystem may contribute to this phenomenon\cite{kimber08}.

	To examine the temperature dependence of the crystal structure and lattice constants and in particular the behavior of the NTE we performed higher resolution neutron powder diffraction at POWGEN for PrFe$_{1-x}$Ru$_x$AsO with x = 0.33, 0.4, 0.5, 0.6 and 0.75.    

	Fig. 6 illustrates a typical refinement using the powder pattern for x = 0.33.  	Table I shows a summary of the lattice and crystallographic parameters obtained at 12 K for several different concentrations of Ru. The angle $\phi$ connecting near-neighbor Fe atoms via an As atom (see Fig. 1) is also included.  To within the resolution of the experiment for each concentration the crystallographic parameters $z_{As}$, $z_{Pr}$, and the angle $\phi$ have only a very small temperature dependence below 200 K. It should be noted however that the precision with which these parameters are determined is an order of magnitude coarser than the precision with which the lattice constants are determined.  This is expected since the lattice constants are determined precisely by the peak positions alone whereas the other parameters are refined by fitting peak intensities.  On the other hand there is a clearly observable dependence of the lattice constants on the temperature over the entire measured range.
\begin{figure}
\includegraphics[width=80mm]{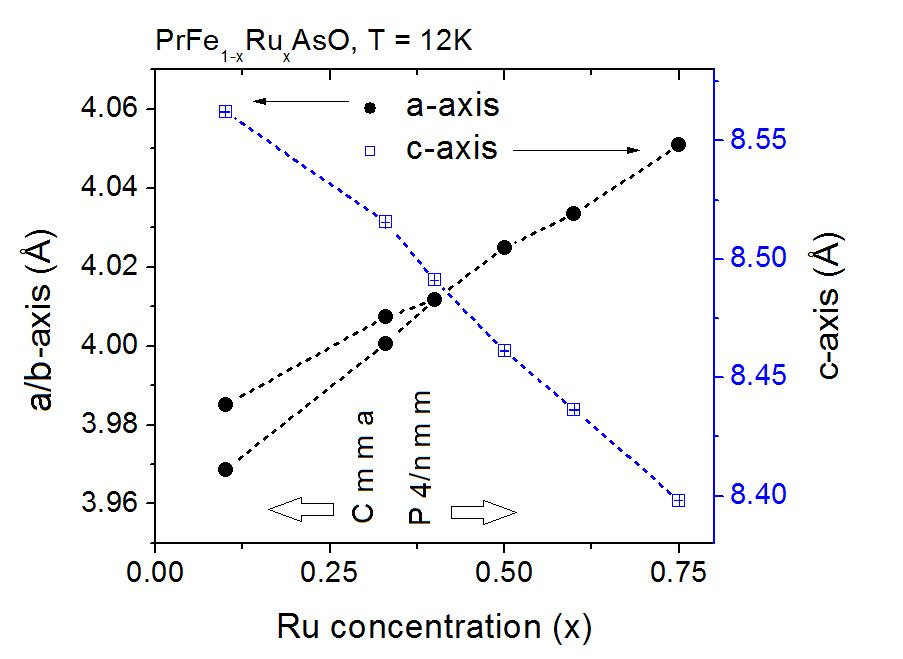}
\includegraphics[width=80mm]{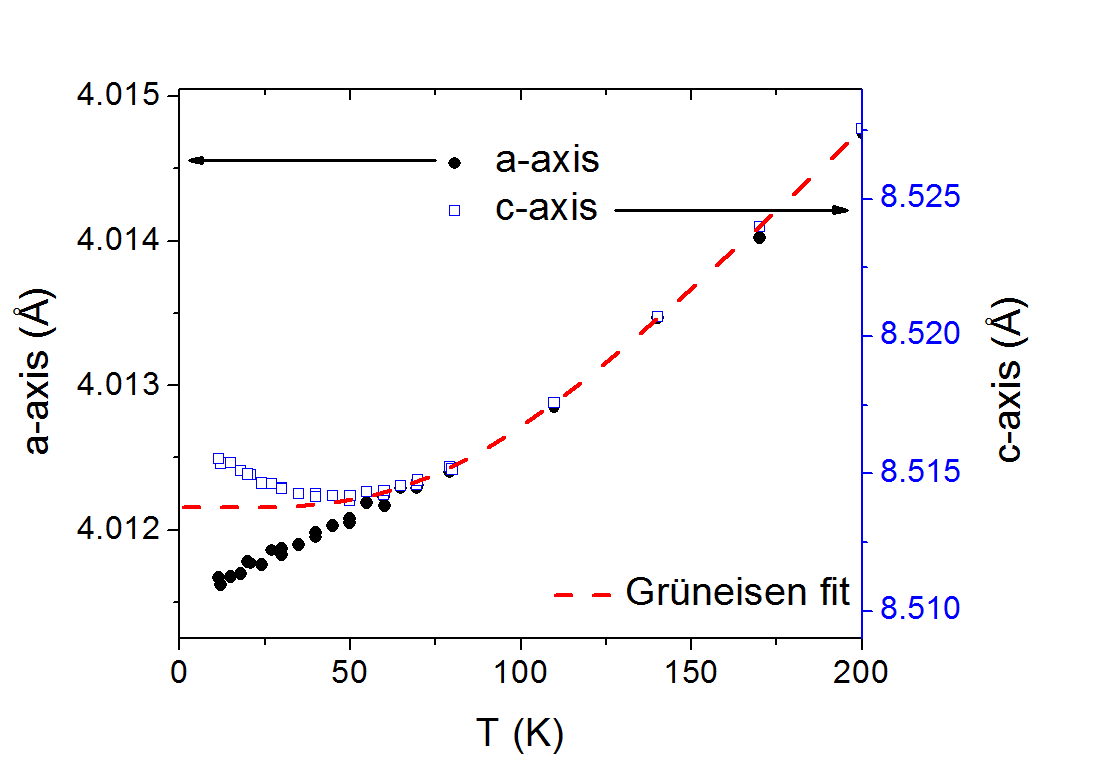}
\caption{(Upper): Doping dependence of a and c at 12 K; (Lower): Temperature dependence of a- and c-axis plotted together, showing the contrasting thermal behavior between the two.  The red line is the Gruneisen model prediction for the c axis based on fits to the data for T $\ge$ 40K as described in the text.} 
\end{figure}
\begin{figure}

\includegraphics[width=80mm]{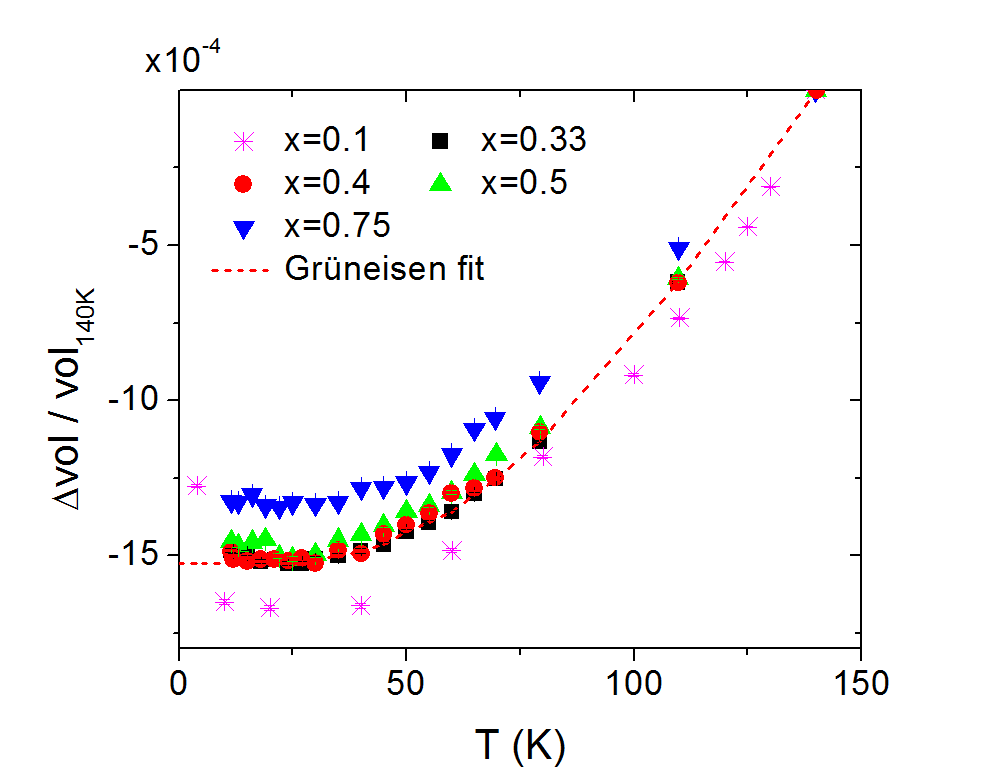}
\caption{Unit cell volume as a function of temperature.  The quantity $\Delta$vol$/vol_{140K}$ is defined as (vol(T)-vol(T=140K))/vol(T=140K). The dotted red line is Gruneisen prediction for $x = 0.33$ fitted using the data for T $\ge$ 40K.} 
\end{figure}

	Fig. 7 (upper) shows the contrast in the behavior of the a- and c-axes' doping dependence at 12K.  The trends are consistent with the room temperature x-ray data reported by McGuire \textit{et al.}\cite{mcguire09a}, which they noted may be related to both the larger radius of Ru compared to Fe as well as constraints on Ru-As and Fe-As bond distances.  Fig. 7 (lower) shows the temperature dependence of the lattice parameters for the concentration x = 0.4. The dashed red line is the extrapolated behavior for the c-axis based on a Gruneisen fit to the volume defined by c$^3$ for the data at 40K and above.  This is a representative concentration as the others show similar behavior. At low temperatures (i.e. for T $\le$ 50K), the lattice parameters deviate from the Gruneisen model fit in opposite directions, somewhat conserving the overall cell volume.  This is illustrated in Fig. 8, which shows the temperature dependence of the unit cell volume. A Gruneisen model fits the entire cell volume better than it fits either the a- or c-axis alone.  This behavior is not uncommon among materials that are anisotropic, where expansion along one direction is often accompanied by contraction along another in order to preserve the overall volume; graphite is one well known example\cite{bailey70}.  Despite this compensation a modest NTE of the unit cell volume is still visible in all samples for T $\le$ 20K.

\begin{figure}
\includegraphics[width=80mm]{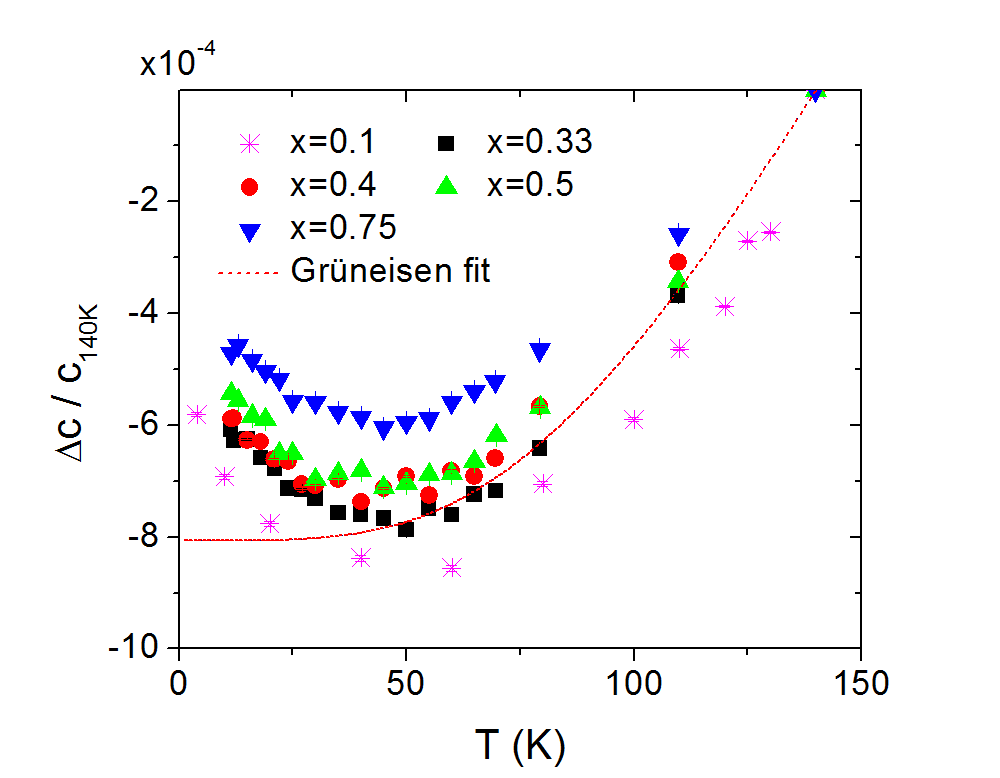}
\includegraphics[width=80mm]{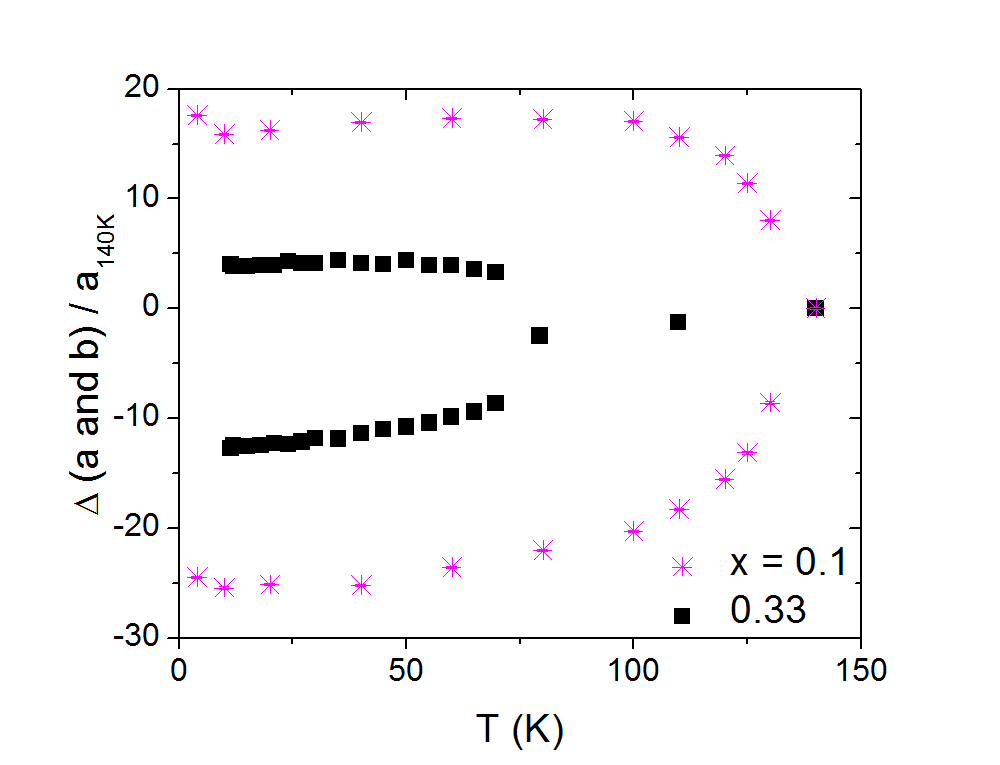}
\includegraphics[width=80mm]{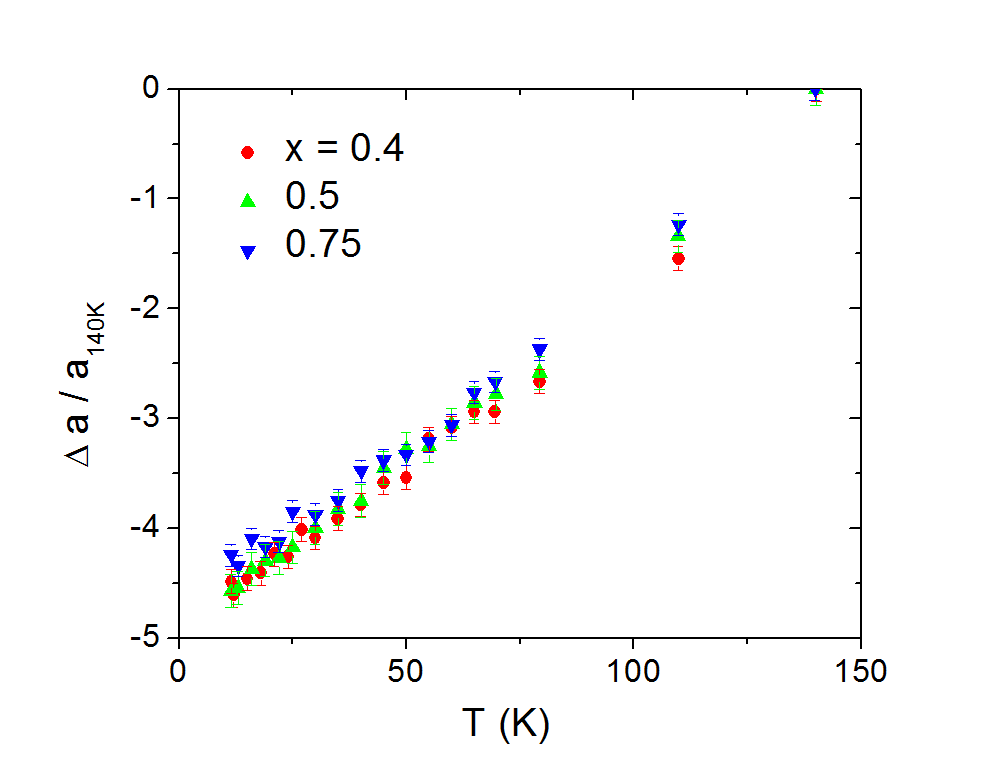}
\caption{ Detailed temperature dependence of of the lattice constants for several values of Ru doping. (Upper): The quantity $\Delta$c$/c_{140K}$ is defined as (c(T)-c(T=140K))/c(T=140K).  NTE is clearly visible in all samples for T $\le$ 50K.  The solid line is a representative Gruneisen model fit for x = 0.33 also shown in Fig. 8. (Middle): A similar plot, showing the relative change in a- and b-axis with respect to lattice parameter a measured at 140K, which is above the structural transition for x = 0.1 and 0.33;(Lower): A similar plot for a-axis for x = 0.4, 0.5 and 0.75, showing the lack of NTE at low temperatures.}
\end{figure}	

	NTE in the c-axis is a prominent feature observed previously in the undoped parent compound: the c-axis reaches a minimum at T = 50K, and then expands as the system is further cooled\cite{kimber08}.  Kimber \textit{et al.} ascribed the presence of c-axis NTE in the non-superconducting parent compound to the effects of spin-lattice coupling\cite{kimber08}.  This was based on their observation that both magnetic order and NTE are absent in the superconducting PrFeAsO$_{1-x}$F$_{x}$, and the proximity of the temperature dependence of the NTE and that of the dome-like feature observed in the intensity of magnetic peaks associated with the ordering of the Fe moments\cite{kimber08}.    In this context, the observation of NTE in PrFe$_{1-x}$Ru$_x$AsO with $x > 0.1$ is surprising, since long range magnetic order is absent.  Fig. 9 shows the temperature dependence of the fractional change in the refined lattice parameters a, b and c for several concentrations.  As seen clearly in Fig. 9 (upper), the anomalous NTE along the c-axis is observed in all samples up to at least x = 0.75.  Figure 10 shows a phase diagram of PrFe$_{1-x}$Ru$_x$AsO with the region of c-axis NTE indicated.

In contrast the thermal behavior of the a-axis is almost linear at low temperatures, with the fractional change almost independent of the Ru concentration.   The existence of c-axis NTE in concentrations without magnetic order suggests that the origin of this behavior might involve more than simply spin-lattice coupling.  Although there is little published data on the c-axis temperature dependence of superconducting 1111 materials, the NTE is absent among those for which data is accessible, including PrFeAsO$_{0.85}$F$_{0.15}$\cite{kimber08}, and LaFeAsO$_{1-x}$F$_{x}$\cite{kondrat09, mcguire08}.  To date we know of no example in the literature of a superconducting 1111 compound that does show NTE.

	It is interesting to compare the low temperature thermal variation of the a- and c-axes with the composition dependence shown in Fig. 7.  In both cases the parameters vary in the opposite direction.  The reason for the opposite temperature dependence of the a- and c-axes remains an open question.  The combination of NTE in the c axis and near conservation of volume may lead to subtle but important changes in the effective magnetic and electronic interactions of the Fe-As layer both internally and with the rest of the sample.   Therefore, it can possibly be inferred that the NTE is linked to the lack of superconductivity (or any new ground state) in the Ru doped samples.

\begin{figure}
\includegraphics[width=85mm]{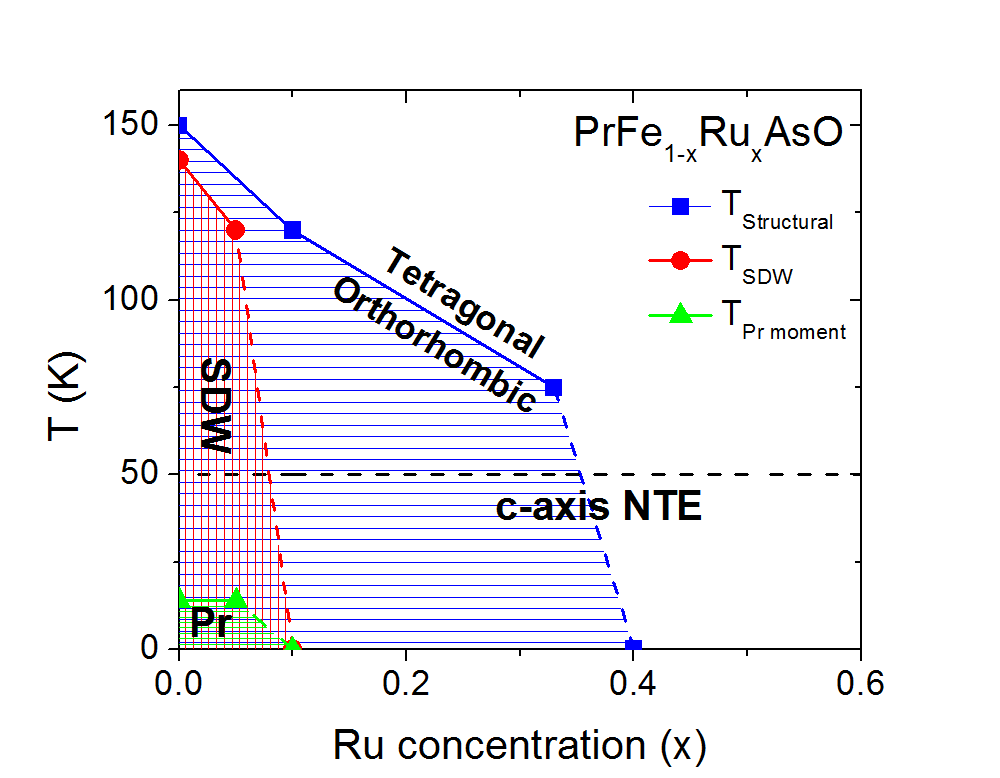}
\caption{Phase diagram of PrFe$_{1-x}$Ru$_x$AsO constructed with data reported in this article, showing temperatures for the tetragonal-orthorhombic transition (T$_{Structural}$), the magnetic Fe ordering (T$_{SDW}$), and the magnetic Pr ordering (T$_{Pr}$), in the measured Ru concentrations.  The region of c-axis NTE presents in all Ru concentrations below T = 50K.}
\end{figure}
\section{\label{sec:level4}Conclusion}
	We have found that magnetic order on both Fe and Pr sites in Pr-1111 is suppressed with 10\% of isoelectronic doping by Ru/Fe substitution, and the tetragonal-orthorhombic structural transition with 40\%.  We have studied the effect of Ru doping on the anomalous NTE along the c-axis and found that the c-axis NTE is present in all of our samples.  For all concentrations the lattice parameter along the c-axis reaches a minimum around 50K.  The NTE is extremely resilient to Ru doping and is measurable up to at least x = 0.75.  The c-axis NTE extends across the entire composition range studied: in compositions which remain tetragonal to the lowest temperatures, those which exhibit tetragonal-orthorhombic distortions with no long-range magnetic ordering, and those that undergo both structural and magnetic phase transitions.  
 	Other anomalous behaviors observed include the near linear temperature expansion of a-axis at low temperature, which also has little discernible effect from doping.  The contrasting thermal behavior between a- and c-axis somewhat conserves the unit cell volume.  The temperature dependence of the unit cell volume agrees better with the Gruneisen prediction than the a- or c-axis alone.  
	It would be interesting to investigate whether the apparent relationship of NTE to the lack of superconductivity is indeed a general rule for 1111 materials.  Such knowledge should provide important insights into the underlying mechanism for superconductivity in Fe based materials.

\section{\label{sec:level4}Acknowledgments}

This research was supported by the US Department of Energy, Office of Basic Energy Sciences.  Experiments were performed at SNS and HFIR, sponsored by the Scientific User Facilities Division.  V. O. G., A. H., and S. E. N. were supported by the Scientific User Facilities Division.  M. A. M. and D. M. were supported by U.S. Department of Energy, Office of Basic Energy Sciences, Materials Sciences and Engineering Division.  Y.Y. was supported by the Office of Basic Sciences, US Department of Energy, through the EPSCoR, Grant No. DE-FG02-08ER46528.

\bibliography{references}

\end{document}